\documentclass[aps,prl,reprint,superscriptaddress]{revtex4-1}

\usepackage{graphicx}
\usepackage{dcolumn}
\usepackage{bm}
\usepackage{subfigure}
\usepackage{float}

\begin{document}

\title{\textbf{\textrm{Coexistence of scale invariant and rhythmic behavior in self-organized criticality}}}

\author{S. Amin Moosavi}
\email{moosavi.s.amin@gmail.com}
\affiliation{Department of Physics, Institute for Advanced Studies in Basic Sciences, Zanjan 45137-66731, Iran}
\affiliation{Department of Physics, College of Sciences, Shiraz University, Shiraz 71946-84795, Iran}
\author{Afshin Montakhab}
\affiliation{Department of Physics, College of Sciences, Shiraz University, Shiraz 71946-84795, Iran}
\author{Alireza Valizadeh}
\affiliation{Department of Physics, Institute for Advanced Studies in Basic Sciences, Zanjan 45137-66731, Iran}

\date{\today}
\begin{abstract}
Scale-free behavior as well as oscillations are frequently
observed in the activity of many natural systems. One important
example is the cortical tissues of mammalian brain where both
phenomena are simultaneously observed.  Rhythmic oscillations as
well as critical (scale-free) dynamics are thought to be
important, but theoretically incompatible, features of a healthy
brain. Motivated by the above, we study the possibility of
coexistence of scale-free avalanches along with rhythmic behavior
within the framework of self-organized criticality. In particular,
we add an oscillatory perturbation to local threshold condition of
the continuous Zhang model and characterize the subsequent
activity of the system.  We observe regular oscillations embedded
in well-defined avalanches which exhibit scale-free size and
duration in line with observed neuronal avalanches.  The average
amplitude of such oscillations are shown to decrease with
increasing frequency consistent with real brain oscillations.
Furthermore, it is shown that optimal amplification of
oscillations occur at the critical point, further providing
evidence for functional advantages of criticality.

\end{abstract}


\maketitle
\section{Introduction}
Generic scale-invariance is ubiquitously observed in natural
systems \cite{Mandelbrot,Kadanoff,B} and has been a subject of
study in physical \cite{P,BTW1,BTW2}, geological \cite{Gutenberg},
biological \cite{BP1} and social sciences \cite{RA}. On the other
hand, rhythmic behaviors such as resonance and synchronization
\cite{Kuramoto} are also observed and studied in many natural
systems \cite{Pikovsky}. From a theoretical point of view these
two phenomena seem to be incompatible, since oscillations imply a
definitive time-scale while scale-free avalanches exhibit no
particular time (or size) scales.

A perfect example of a system where both phenomena are observed is
the collective neural dynamics of mammalian cortex. On one hand,
rhythmic oscillations of cortical neurons are well documented and
intensively studied in regard to their formation
\cite{Buzsaki,WB,Wang2,Valizadeh} as well as their functional and
behavioral correlates \cite{Wang}. On the other hand, neuronal
avalanches \cite{BP1,BP2,TBFC,FIBSLD,PTLNCP,SACHHSCBP,C} with
scale-free statistics of their size and duration imply lack of
time and size scales for brain dynamics. It has been shown that
coexistence of these two phenomena is important for development of
cortical layers \cite{Elakat}. Scale-free behavior of the brain,
that is thought to be the result of underlying criticality, has
recently attracted much attention in regard to optimum dynamic
range in response to stimulus \cite{SYPRP,KC}, functional
robustness \cite{NIK}, learning capability \cite{AH}, information
processing \cite{Beggs} and transmission \cite{SYYRP}.

The important issue here is how these two phenomena, that seem to
be incompatible in a first pass, emerge simultaneously in the
cortex. Moreover, what would be the consequences of such a
coexistence. Despite the importance of this phenomena, a few
theoretical studies  have been devoted to this subject. Poil et
al. have shown that this phenomena can emerge as a result of a
balance between inhibition and excitation \cite{Poil}. More
recently Wang et al. have shown that this coexistence can emerge
as a finite-size effect in self-organized critical states of small
systems \cite{Wangg}. Emergent stochastic oscillations are also observed close to the critical point in non-conservative systems of interacting excitable nodes known as self-organized quasi-critical models \cite{Costa,CBCC}.

The typical theoretical framework for the study of rhythms of the
brain is synchronization of coupled phase oscillators such as the
Kuramoto model \cite{Kuramoto} while neuronal avalanches are
mostly studied in the context of self-organized criticality (SOC)
\cite{LHG,Osame,Hesse,Wangg,AMIN,AMIN1,Munoz} or models of
excitable nodes \cite{KC,Larremore,Levina1,Poil,Amin3}. In this
work, we intend to study the effects of oscillations in the
framework of SOC. In particular, we introduce and justify an
oscillatory perturbation into a well-known Zhang model of SOC
\cite{Zhang,SV} and subsequently characterize its response to such
a perturbation. Interestingly, we find that oscillations
dominantly occur while embedded within well defined avalanches
which exhibit scale-free statistics for their size and duration.
We further find that the well established response of the system
is further enhanced and amplified at the critical point leading to
large amplitude oscillatory behavior as a result of subthreshold
oscillatory perturbations.

\section{THE MODEL}

In order to study the behavior of a self-organized critical (SOC)
model under the influence of an external oscillatory perturbation
we use a sandpile model known as the stochastic Zhang sandpile
model \cite{SV}. The reasons for choosing this model for our study
is that it exhibits continuous dynamical variables, a threshold
dynamics that can mimic the neuronal dynamics and well-behaved
scaling behavior \cite{abdolvand}. The model is considered on a
two dimensional $L\times L$ square lattice (number of sites in the
lattice is $N=L^2$) with nearest neighbor interactions. Every node
of the lattice $i$ is assigned a dynamical variable $V_{i}$ (e.g.
energy or potential). Dynamics of the model exhibits a slow perturbative drive
during which small amounts of energy are added to the system, i.e.
$V_{i}=V_{i}+\delta V$ where $\delta V$ is a randomly chosen real
number from $[0,0.25]$. This process is continued until a site
reaches a threshold value ($V_{th}=1$) at this point a fast
dynamics takes place. Addition of energy via perturbative drive is
not possible during an avalanche in SOC systems. This property is
known as separation of slow and fast time scales
\cite{P,BTW1,BTW2}. The rule for the fast dynamics is that each
site $i$ with $V_{i}\geq V_{th}$ becomes unstable and distributes
its energy between its neighbors with the following toppling rule
\begin{equation}
      \label{EQ1}
      \begin{array}{l}
            if \hskip15pt V_{i}>V_{th}\\
            then \hskip5pt V_{j}\rightarrow V_{j}+W_{j}V_{i} ,\hskip5pt V_{i}\rightarrow 0
       \end{array}
\end{equation}
in which the index $j$ is related to all neighbors of site $i$,
and $W_{j}$ are annealed random numbers in the range $[0,1]$ with
the constraint of $\sum_{j}W_{j}=1$ which leads to strict local
conservation of energy.  Conservative dynamics in addition to
separation of time-scales are believed to be required for
observation of self-organized criticality. However, it has been
shown that breaking local conservation does not violate
criticality of the system as long as the dynamics is on the
average (or globally) conservative \cite{AMIN,AMIN1}. It
is generally believed that neuronal interactions are not
conservative where different electrophysiological mechanisms play
role in delivering electrical signal via synapses. Therefore,
models of non-conservative interacting neurons have been developed
as self-organized quasi-critical models where approximate
criticality is considered as an explanation of scale-free behavior
of neuronal avalanches \cite{LHG,Bonachela,Costa,CBCC}. However,
in this work we use a conservative SOC model which can be
considered as a limiting case of a more realistic model of
neuronal dynamics.

The fast dynamics pursuing Eq. (\ref{EQ1}) is triggered by a single
toppling. As a result of a toppling it is possible that the
neighbors that receive energy become unstable and topple in the
next time step and a cascade of toppling takes place which is
called an avalanche. An avalanche ends when there are no unstable
sites in the system. Here we must note that all the sites that
become unstable in one time step will topple together, i.e. we use
the parallel update rule \cite{foot}. Boundaries of the lattice
are open and energy can be dissipated through the boundaries. Size
($S$) of an avalanche is defined as the number of topplings and
duration ($D$) of an avalanche is defined as the number of time
steps (parallel updates) of the avalanche.

We also use another version of sandpiles known as fixed energy
sandpile model \cite{Montakhab,Vespignani}. In this version,
periodic boundary conditions are imposed on the system and the
external perturbation is turned off. Therefore, the average energy
of the system ($E=1/N \sum_{i=1}^{N}V_{i}$) is fixed by the
initial conditions. Fixed energy sandpile models exhibit a control
parameter which is the average energy of the system $E$ and an
order parameter which is the long term average of density of
active nodes $\rho$ in the system. Activity is initiated by
choosing a random site $i$ and allowing it to topple according to
Eq.(\ref{EQ1}), regardless of its value $V_i$. This model exhibits
a continuous (dynamical) phase transition, passing through a
critical point $E=E_{c}$, from an absorbing state where any
activity ends ($\rho=0$) to a running state where one observes
ceaseless dynamics ($\rho>0$). Properties of the system at the
critical point of fixed-energy sandpile is in accordance with its
SOC counterpart \cite{Montakhab}. Using fixed-energy sandpiles, we
can study the behavior of the system in the sub-critical
($E<E_{c}$) as well as super-critical ($E>E_{c}$) phases.

We now introduce an oscillatory perturbation to the system. During
the fast dynamics we simply introduce a sub-threshold oscillatory
perturbation to the model that changes the dynamics by changing
the condition of toppling as follows:
\begin{equation}\
\label{EQ2}
      \begin{array}{l}
            if \hskip15pt V_{i}+\delta \times f(\Omega t+\phi_{0})>V_{th}\\
            then \hskip5pt V_{j}\rightarrow V_{j}+W_{j}V_{i} ,\hskip5pt V_{i}\rightarrow 0
       \end{array}
\end{equation}
in which $f$ is a normalized oscillatory function, $\delta$ is the
(sub-threshold) amplitude of the oscillatory perturbation,
$\Omega=2\pi/T$ is the angular frequency of the oscillations ($T$
is the duration of oscillatory perturbation) and $\phi_{0}$ is the
initial phase that is chosen randomly from $[0,2\pi]$ at the
beginning of each avalanche. Here, we must note that the
oscillatory perturbation does not add energy to the system it just
manipulates the condition of toppling and the dynamics is,
regardless of the oscillations, strictly conservative. In this
paper we will show that it is possible to introduce a time-scale
to the dynamics of a critical system while it still remains at the
critical point.

An example of a physical situation for our model is a cortical
tissue that receives sub-threshold oscillatory input from any
other parts of the brain. A real neuron undergoes oscillations in
its membrane potential when receiving a sub-threshold oscillatory
input current \cite{Gerstner}. Subsequently, the excitability of a
neuron becomes an oscillatory function of time. Therefore, our
model captures in a simple way a threshold and release mechanism,
driven by an external oscillatory plus local inputs, and thus
resembles what one expects from real neuronal dynamics.

\section{RESULTS}

 \begin{figure*}[!htbp]
\includegraphics[width=0.99\linewidth]{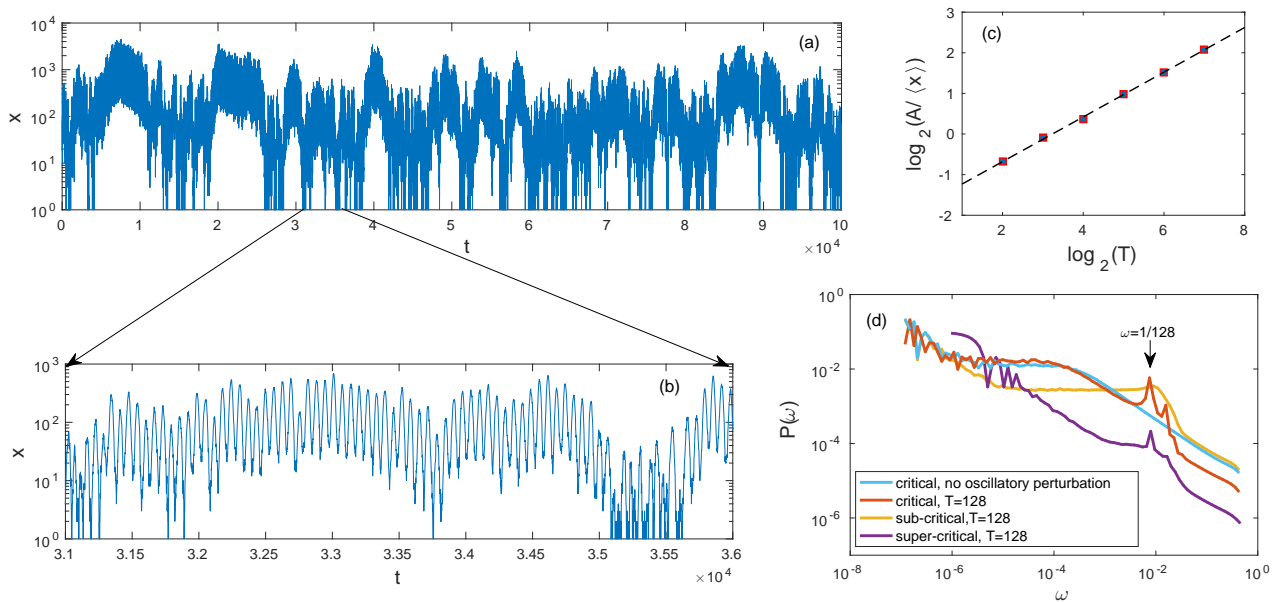}
\caption{(Color on line)(a) Activity of the system (consecutive
avalanches) as a function of time for a system with $T=64$. An
avalanche starts when $x>0$ and ends when $x=0$. Note the
logarithmic scale on the vertical axis. (b) Oscillations of the
activity embedded in avalanches. (c) Normalized amplitude of
oscillations in activity as a function of period of oscillations.
(d) Power spectrum of $x$ for different phases of systems with
oscillatory perturbations with $T=128$ at($E=E_{c}=0.596$), above
($E=0.608$) and bellow ($E=0.585$) the critical point, compared
with the critical system without oscillatory perturbation. We set
$L=1024$, $f=\sin(\Omega t+\phi_{0})$ and $\delta=0.1$ in all
panels.} \label{FIG1}
\end{figure*}

To understand the scaling as well as oscillatory behavior of the
system, we performed extensive computer code simulations of the
systems with different sizes of $L=512,1024,2048$. First, we focus
on the activity of the system ($x$) which is defined as the number
of active sites (the sites that topple) at each time step of an
avalanche. An avalanche starts when $x$ becomes equal to one and
ends when it becomes zero. We find that systems that are
influenced by the oscillatory perturbation exhibits oscillations
in activity embedded in avalanches, i.e. during each avalanche $x$
is an oscillatory function of time with a period equal to the
oscillatory perturbation (see Fig.\ref{FIG1}(a,b)).

In order to study the properties of oscillatory perturbations we calculate the average amplitude of oscillations as
\begin{equation}
\label{EQ3}
 A=\langle A_{k}\rangle=\Bigg\langle \langle x^{k}_{max}\rangle-\langle x^{k}_{min}\rangle \Bigg\rangle
\end{equation}
where $\langle x^{k}_{max}\rangle$ and $\langle x^{k}_{min}
\rangle$ are respectively the average of maxima and minima of
oscillatory activity in an oscillatory avalanche $k$. Large
brackets are for averaging over all oscillatory avalanches. An
interesting property of these oscillations is that their
normalized amplitude increases as a power-law function ($A/\langle
x \rangle\sim T^{0.55}$ where $\langle x\rangle$ is the average
activity of the system over all oscillatory avalanches) of the
period of oscillatory perturbation (see Fig.\ref{FIG1}(c)). This
property is in agreement with our knowledge of the rhythms of the
brain \cite{Buzsaki} where we observe low amplitude, high
frequency oscillations and vice versa. For example, low frequency
alpha ($\sim 10$Hz) oscillations occur at relatively high
amplitude while high frequency gamma ($30-80$Hz) oscillations
occur at low amplitudes in the cortex.

Power spectrum of the activity of system is calculated and plotted
in Fig.\ref{FIG1}(d) for different fixed-energy models with and
without oscillatory perturbations at $E=E_{c}$, and also above and
bellow the critical point. In the sub-critical phase we observe a
flat line for low frequencies which is an indication of a noisy
dynamics. Comparing to the critical state of the system without
oscillatory perturbation, it is clear that for the critical system
with oscillatory perturbation, one obtains a power-law function
with a peak at the frequency of oscillatory perturbation. This
behavior is ubiquitously observed in electroencephalogram as well
as local field potential analysis of many parts of the brain
\cite{Biyu}. We have done the same analysis for different values
of $T=128,64,32$ and we observed the same behavior in all cases.
Therefore, we can conclude that the system exhibits a wide range
of frequencies that have the potential to be amplified and
observed at the critical point. A degree of amplification of
oscillatory perturbations is observed in the super critical phase
with an amplitude smaller than the the observed amplitude in the
critical phase (see Fig.\ref{FIG1}(d)). Using a quantitative
analysis, we will show later that the amplification is maximized
at the critical point compared to off-critical phases. We note
that in contrast to the phenomenon of resonance where
amplification of special frequencies is possible, this
amplification is possible for all the power-law distributed
frequency range. This behavior is in line with oscillatory
behavior in the brain where a wide range of frequencies is
observed \cite{Buzsaki,Biyu}.

\begin{figure*}[!htbp]
\includegraphics[width=0.99\linewidth]{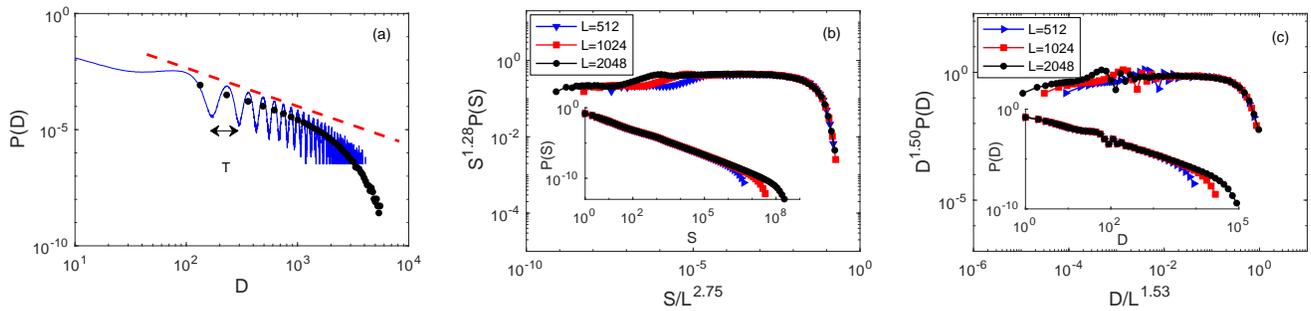}
\caption{(Color on line) (a) Probability distribution function of
duration of avalanches ($P(D)$) is plotted for $L=512$,
$f=\sin(\Omega t+\phi_{0})$, $\delta=0.1$ and $T=128$. Circles are
the results of averaging over intervals of $\Delta D=T$ which
exhibit power-law behavior shown by the dashed line. (b,c) Main
panels are finite-size scaling collapse of rescaled data into one
universal curve for probability distribution functions of duration
($D$) and size ($S$) of avalanches where $\tau_{S}=1.28$ and
$\tau_{D}=1.50$. Insets are plots of probability distribution
functions of $S$ and $D$ corresponding to the main panels.}
\label{FIG2}
\end{figure*}

In order to study the scaling behavior of the system we focus on
avalanche statistics. A prototypical example of probability
distribution function of duration of avalanches ($P(D)$) is
plotted in Fig.\ref{FIG2}(a) for $L=512$, $f=\sin(\Omega
t+\phi_{0})$, $\delta=0.1$ and
$T=128$. It is interesting that $P(D)$ exhibits a bouncing
behavior for large enough $D$ over intervals of $\Delta D=T$. And,
if we average the data over time bins of $\Delta D=T$, we observe
power-law behavior of the binned data. Notably, this bouncing
behavior is not observed for the probability distribution function
of avalanche sizes and a standard power-law behavior is observed
for $P(S)$.

\begin{figure*}[!htbp]
\centering
\includegraphics[width=0.99\linewidth]{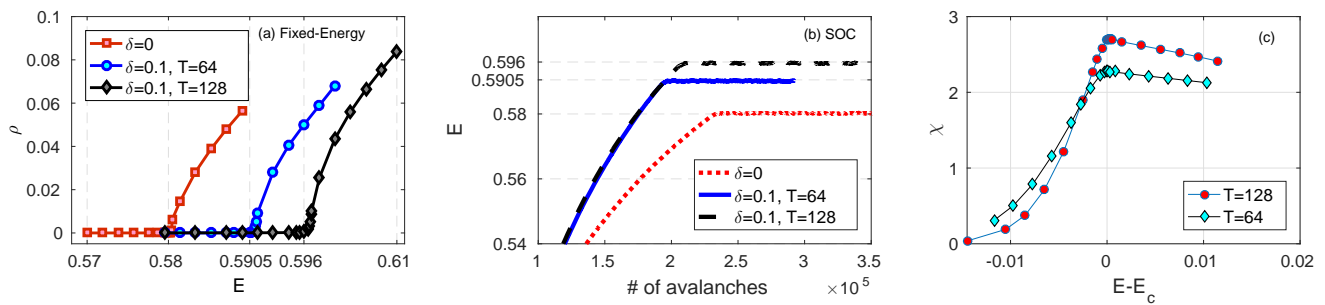}
\caption{(Color on line) (a) Second order (continuous) phase
transition from absorbing to running state for fixed energy
sandpile models with oscillatory perturbations. (b) Self-organization of average energy where after a transient time the open system with external drive settles into the critical steady state in
agreement with the transition point of the fixed-energy model shown in panel (a). (c) Average
amplitude of oscillations as a function of the control parameter
$E$. } \label{FIG3}
\end{figure*}

Simply observing extended scaling for a finite system is not
necessarily proof of criticality.  To verify establishment of
self-organized criticality and also evaluating scaling exponents,
we perform a finite-size scaling of our data for different system
sizes of $L=512,1024,2048$. We consider a simple scaling ansatz
for the probability distribution function of size and duration of
avalanches:
\begin{equation}
      \label{EQ4}
     P(y)\sim y^{-\tau_{y}}G_{y}(y/L^{\beta_{y}})
\end{equation}
where $y$ can be either the binned data of $D$ or $S$, $\tau_{y}$
is the critical exponent, $\beta_{y}$ is the finite size exponent
determining the cutoff in $P(y)$, and $G_{y}$ is the universal
function that, in the case of criticality, exhibits the same shape
for all system sizes \cite{P}. If the system is critical, and we
rescale $y\rightarrow y/L^{\beta_{y}}$ and $P(y)\rightarrow
y^{\tau_{y}}P(y)$, then the plots of rescaled data must collapse
into one universal curve for different system sizes. In
Fig.\ref{FIG2} (c,d) we present the results of finite-size scaling
analysis of our data. We observe good collapse of data for both
cases of $D$ and $S$ with $\tau_{S}=1.28(1)$, $\tau_{D}=1.50(1)$,
$\beta_{S}=2.75(1)$ and $\beta_{D}=1.50(1)$. These values of
exponents are in agreement with the scaling exponents in the
absence of oscillatory perturbations \cite{P,AMIN,Vespignani},
which indicates robustness of scaling behavior of SOC systems in
the presence of oscillatory perturbations. Here, we can conclude
that despite having oscillatory behavior the system exhibits scale
invariance and criticality. We note that the same analysis were
performed for different $f=\sin(\Omega t+\phi_{0}),f=\sin(\Omega
t+\phi_{0})+1,f=\sin(\Omega t+\phi_{0})-1$ as well as different
values of $\delta=0.5,0.1,0.2$ and the same behavior is observed
in all cases.

To better understand the behavior of the system in sub- and
super-critical phases we performed computer simulations of the
above explained fixed-energy sandpile model with $L=1024$,
$T=64,128$ and $\delta=0.0,0.1$. Fig.\ref{FIG3}(a) is a plot of
the order parameter $\rho$ versus the control parameter $E$. It is
clearly seen that there is a continuous phase transition at the
critical point $E_c$ which is in agreement with the $E_c$ that is
obtained from the self-organization process in the standard form
of the model with open boundary conditions and external drive (see
Fig.\ref{FIG3}(b)).

The key feature of neural networks poised close to a standard
second order phase transitions is \lq\lq optimum" response to
stimulus \cite{KC,Beggs}. So far, we have essentially added small
subthreshold oscillations to a well-known model of SOC, and have
characterized their effect on the activity of the system. We next
ask to what extend criticality provides amplifications of such
oscillations. In order to quantify the amplification of
oscillatory perturbations, we define $\chi$ as the expectation
value of average amplitude of oscillations over all active times
\begin{equation}
\label{EQ5} \chi=\Big\langle P_{k}\frac{A_{k}}{\langle x
\rangle}_{k}\Big\rangle
\end{equation}
where $P_{k}$ is the probability of having a rhythmic behavior in
an avalanche $k$, which is a binary probability, i.e. it is equal
to one if there is an oscillation in the avalanche and is equal to
zero otherwise. $A_{k}$ is the average amplitude of oscillations
as defined in Eq.(\ref{EQ3}) and $\langle
x\rangle_{k}=S_{k}/D_{k}$, is the average activity during the
$k^{th}$ avalanche, which serves as normalization. Large brackets
indicate averaging over all avalanches. Fig.\ref{FIG3}(c) shows a
plot of $\chi$ around the critical points as a function of
$E-E_{c}$ for $T=64,128$. It is interesting that amplification of
oscillations is maximized at the critical point regardless of the
value of $T$. However, we observe larger amplification for slower
oscillations around the critical point which is in agreement with
our results of Fig \ref{FIG1}(c). We therefore conclude that
optimal amplification of sub-threshold oscillations occurs at
criticality. This has important consequences for brain function as
production of rhythms are thought to be key elements in coding and
transfer of information in the brain. This yet provides another
motivation for the \emph{critical brain hypothesis. }

\section{CONCLUDING REMARKS}

Motivated by critical as well as oscillatory dynamics of
neuro-cortical circuits, we have analyzed a simple model of SOC
which is influenced by sub-threshold oscillations. Interestingly,
we find that the system exhibits well-defined oscillations
embedded in avalanches where the average amplitude of oscillations
is an increasing power-law function of the period of oscillations.
Consequently, an off-critical system, that exhibits a time scale
$D_{max}$ for avalanches, cannot respond to a wide range of
frequencies and is thus limited in range of oscillatory activity
it can exhibit. However, due to scale-free behavior of avalanches
at the critical point one observes a proper response to all
frequencies. This could be important for a functional brain since
we observe a wide range of frequency of rhythms in different
regions of the brain.

One might be tempted to associate the observed amplification of
system's response to stochastic resonance (SR), as there too, one
observes the amplification of external (subthreshold) drive
frequency in a stochastic background. However, the mechanisms are
entirely different. The criticality associated with the collective
dynamics of our model is capable of amplifying a wide range of
subthreshold frequencies, without need to tune any system
parameter. On the other hand, in SR one needs to tune the noise
intensity (and thus system's natural frequency) in order to see
amplification in response for a given frequency.

Our finite size scaling analysis of the
statistics of size and duration of avalanches suggests that, despite
having oscillations, the system exhibits critical properties in
agreement with the systems without oscillatory perturbations. We showed that the same exponents are observed with oscillatory perturbation and thus the
robustness of criticality as well as universality class was
confirmed.  We note that divergence of avalanche durations with
system size implies that duration of spontaneous oscillations in
the resting state of the brain should only be bounded by the size
of the cortex.

We also note that the exponents we obtained for the Zhang model
are not the same as the standard mean-field exponents for real
neuronal avalanches. However, on can imagine structural as well as
dynamical modifications to our model which could lead to
mean-field exponents. For example, our $2D$ nearest neighbor
interaction is not a good topology for the real cortex.  Larger
average connectivity along with random neighbor would lead to mean
field behavior, which is the exact solution for an all-to-all
network model. Furthermore, as has been shown in Ref.~\cite{AMIN},
addition of synaptic noise in the dynamics could also lead to mean
field behavior independent of the structure of the network chosen.

It has been shown that criticality of the brain leads to many
advantages for the brain functions
\cite{SYPRP,KC,NIK,AH,Beggs,SYYRP}. Due to the crucial role of
oscillations in brain functions, it is very important that the
brain respond to oscillatory perturbations efficiently. We show
that the optimum amplification of oscillatory perturbations takes
place at the critical point. In other words, not only the system
remains critical but also amplification of oscillations is allowed
over a wide range of frequencies. This optimum amplification can
be the root to optimum signal coding and transmission by
oscillations over different time and length scales.

\begin{acknowledgments}
Support of Iranian National Elites Foundation, Institute for
Advanced Studies in Basic Sciences, and Shiraz University Research
Council is acknowledged. S.A.M. also acknowledges Y. Sobouti
without whose support this work could not be completed.
\end{acknowledgments}
\bibliographystyle{apsrev}
\bibliography{xbib}

\begin{thebibliography}{99}
\bibitem{foot} {A sequential update rule is also
possible. Differences between parallel and sequential update rules
and their effects on the critical properties of sandpile models
are discussed in references \cite{P,SV}.}
\end{thebibliography}

\end{document}